# Image Encryption and Decryption in A Modification of ElGamal Cryptosystem in MATLAB

Hayder Raheem Hashim[a*], Irtifaa Abdalkadum Neamaa[b]

[a] *Department of Mathematics, Faculty of Computer Science and Mathematics, University of Kufa, Najaf, Iraq*
*hayderr.almuswi@uokufa.edu.iq*

[b] *Department of Mathematics, Faculty of Computer Science and Mathematics, University of Kufa, Najaf, Iraq*
*irtefaa.radhi@uokufa.edu.iq*

## Abstract

The need of exchanging messages and images secretly over unsecure networks promoted the creation of cryptosystems to enable receivers to interpret the exchanged information. In this paper, a particular public key cryptosystem called the ElGamal Cryptosystem is presented considered with the help MATLAB Program to be used over Images. Since the ElGamal cryptosystem over a primitive root of a large prime is used in messages encryption in the free GNU Privacy Guard software, recent versions of Pretty Good Privacy (PGP), and other cryptosystems. This paper shows a modification of the this cryptosystem by applying it over gray and color images. That would be by transforming an image into its corresponding matrix using MATLAB Program, then applying the encryption and decryption algorithms over it. Actually, this modification gives one of the best image encryptions that have been used since the encryption procedure over any image goes smoothly and transfers the original image to completely undefined image which makes this cryptosystem is really secure and successful over image encryption. As well as, the decryption procedure of the encrypted image works very well since it transfers undefined image to its original. Therefore, this new modification can make the cryptosystem of images more immune against some future attacks since breaking this cryptosystem depends on solving the discrete logarithm problem which is really impossible with large prime numbers .

------------------------------------------------------------------------

* Corresponding author.
E-mail address: hayderr.almuswi@uokufa.edu.iq.

## 1. Introduction

In cryptography, the encryption process is the process of transforming data using an algorithm or a mathematical function to make it unreadable to anyone except the intended receiver who knows the private key. The ElGamal cryptosystem is a public key cryptosystem technique, whose security is based on the difficulty solving the discrete logarithm problem [1]. This cryptosystem was invented in 1985 by the Egyptian cryptologist Taher ElGamal. However this cryptosystem works over the finite cyclic group $Z^*_p$ of the finite field [2], it's also used over a primitive root of a large prime number. However the ElGamal cryptosystem is also used in encrypting and decrypting texts, e-mails, files, and software, it is not obvious to be used in encrypting and decrypting images such that gray and color images with help of MATLAB program. In imaging science, image processing is any form of signal processing for which the input is an image, such as a photograph or video frame; the output of image processing may be either an image or a set of characteristics or parameters related to the image. Most image-processing techniques involve treating the image as a two-dimensional signal and applying standard signal-processing techniques to it. Also, image processing refers to digital image processing, but optical and analog image processing also are possible [8].

*The Types of Digital Gray and Color Images:*

• Gray scale: Each pixel is a shade of gray, normally from 0 (black) to 255 (white). This range means that each pixel can be represented by eight bits, or exactly one byte. Other grey scale ranges are used, but generally they are a power of 2.

• Color :True Color, or RGB: Each pixel has a particular color; that color is described by the amount of red, green and blue in it. If each of these components has a range 0–255, this gives a total of 2563 different possible colors. Such an image is a "stack" of three matrices; representing the red, green and blue values for each pixel. This means that for every pixel there correspond 3 values [3]. The cryptography in digital computing has been applied to different types of digital file formats such as text, images video etc.Indeed, because the need of information security, image encryption and decryption has been an important research area and it has wide application prospects. Therefore, this article focuses on applying one of the most public key cryptosystems ,which is the ElGamal cryptosystem over a primitive root of a large prime number, over images using MATLAB.

## 2. Primitive Root
### 2.1. Definition

Let n and r are relatively prime integers with n>0 .If $r^x \equiv 1 \pmod{n}$ holds when ø(n) = x , where ø(n) is the Euler's phi-function, then r is called a primitive modulo n [5]. If n=p which is a prime number that is satisfied the definition of the primitive root ,then r is a primitive root of p. In addition, In 1769 Lagrange proved that every prime has a primitive root [4].

*2.2. Example*

The integer r =3 is a primitive root of the prime number p=7 , since $3^6 \equiv 1 \pmod{7}$ holds if $\emptyset(7) = 6$.

3. **The ElGamal Cryptosystem**

The ElGamal cryptosystem is a well- known cryptosystem, invented by T.ElGamal in 1985. Its security is based on the difficulty of finding discrete logarithms modulo a large prime. In the ElGamal cryptosystem, each person choose a very large prime number p, a primitive root r of p, and an integer a with $2 \leq a \leq p-2$. This integer a is the private key that must be kept secret by that person, and the corresponding public key is (r, s, p) such that, $s \equiv r^a \pmod{p}$. The message M can be encrypted to the pair (x,y) such that $x \equiv r^k \pmod{p}$ and $y \equiv (m * s^k) \pmod{p}$. Then encrypted message (x,y) can be decrypted by $M \equiv [y((x)^a)^{-1}] \pmod{p}$. Breaking this cryptosystem depends on finding a which is a unsolvable conjuncture in mathematics called the discrete logarithm problem, because it needs thousands of years to find all the possible solutions of it. The ElGamal cryptosystem is well-known to be used in encrypting and decrypting texts, e-mails, files, and software [4]. Therefore the following is a modification of using the ElGamal cryptosystem over a primitive root of a large prime in encrypting and decrypting gray and color images with help of MATLAB program.

*3.1. Image Encryption*

Image encryption techniques try to convert an image to another image that is hard to understand; to keep the image confidential between users, in other word, it is essential that nobody could get to know the content without a key for decryption [7]. All the public key cryptosystems have a public and a private key. The public key is used in the encryption procedure and can be published, while the private key must be possessed only by the recipient of the message and used in the decryption procedure [6].

*The key-generation process is as the following:*

- Select a large prime p and a positive integer r such that r is a primitive root of p.
- Select a positive integer a such that $1 < a \leq p-2$.
- Compute $s \equiv (r)^a \pmod{p}$.
- Let the public-key be announced publicly). →(r, s, p).
- Let the private key be kept secret →(a)

To encrypt an image (gray, color) only the public key (r, s, p) is needed, thus a cipher image(encrypted image) is produced out of the plain image (original image).

*The encryption process Algorithm is as the following ( done by MATLAB)*

- Step1: start
- Step2:Read the plain image into its corresponding matrix (call it M ) using MATLAB such that each

element $m_{ij}$ in M does not exceed the prime number p.

- Step3 : Let the public key (r, s, p) .
- Step 4: For all the element $m_{ij}$ in the matrix, select one random integer k with , $1 \leq k \leq p-2$ .
- Step 5: Compute $X \equiv r^k$ (mod p).
- Step 6: Compute $y_{ij} \equiv (m_{ij} * s^k)$ (mod p)
- Step 7: Show the encrypted image Y .

Now, after reading the encrypted image, the matrix of the encrypted image can be sent as Y. In addition, the sender has to send X with Y to the receiver. Once the encrypted image reaches the destination, it must be decrypted .

*3.2. Image Decryption*

To decrypt the cipher image, the private key **a** and **X** are necessary to be known by the receiver. The process as the following (done by MATLAB):

- Step1: Import to data (Y) from the encrypted image .
- Step 2: Restore the plain image M, such that $M \equiv [Y((X)^a)^{-1}]$ (mod p)
- Step3: Obtain the original image(decrypted image).

The following flow chart shows the image Encryption and Decryption process :

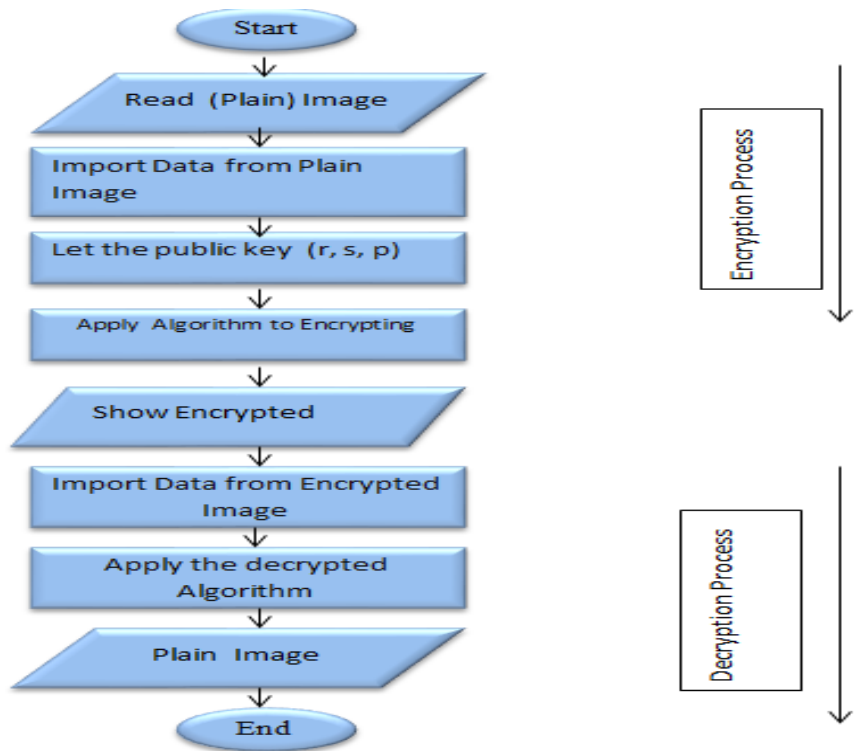

*Fig(1):The Flow Chart Diagram for Encryption/Decryption Image*

## 4. Simulation And Result

In this paper we have simulated the image processing part of encryption and decryption in MATLAB software. First of all we would take an image and then we obtain its corresponding matrix. Then we would be encrypting the image matrix using a modification the ElGamal cryptosystem . The result shows the original images (color and grayscale ) and encrypted images . Note that the decrypted image is the same as the original image .

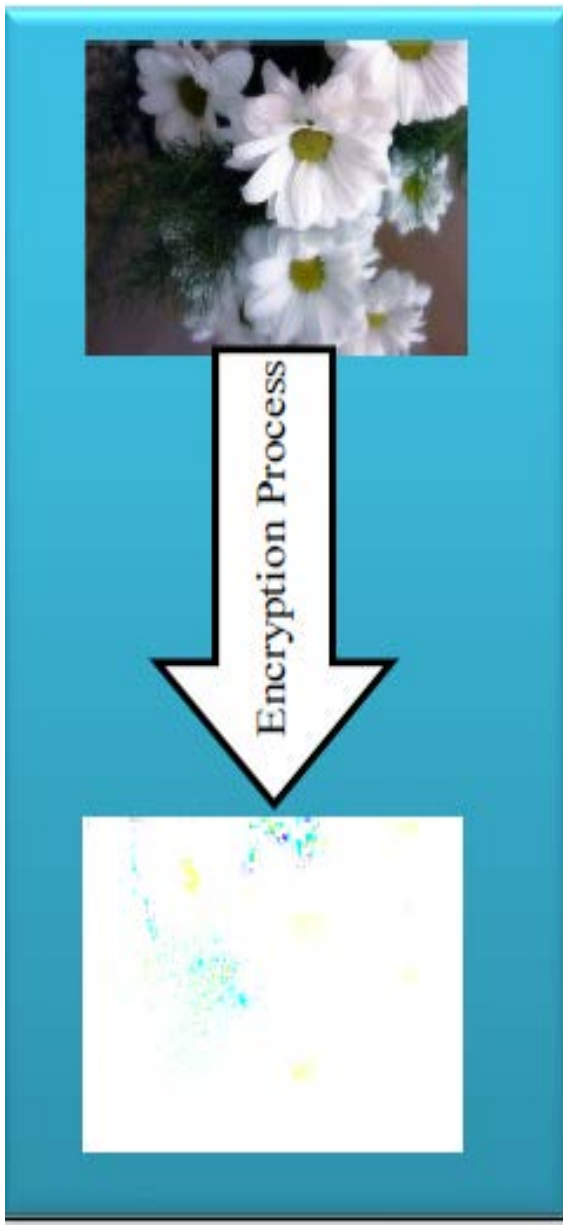
Fig.(2): Encryption of Color Image (jpg)

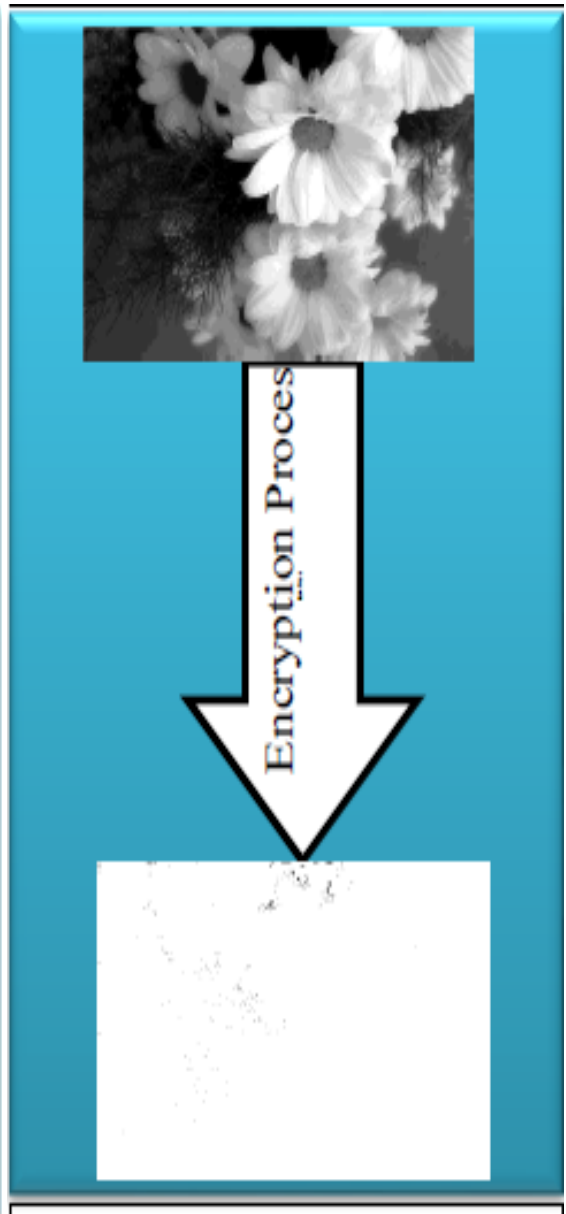
Fig.(3): Encryption of Gray Image (jpg)

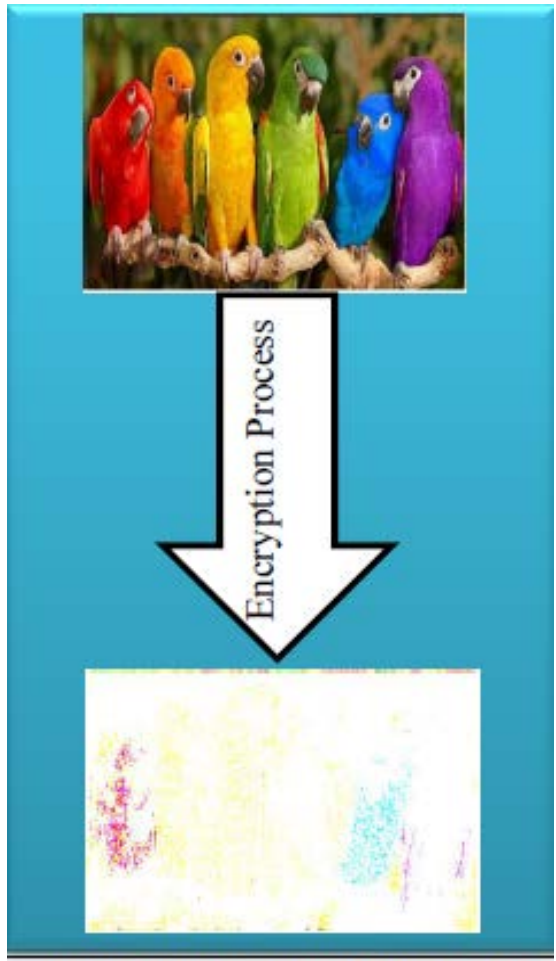

Fig.(4): Encryption of Color Image (png)

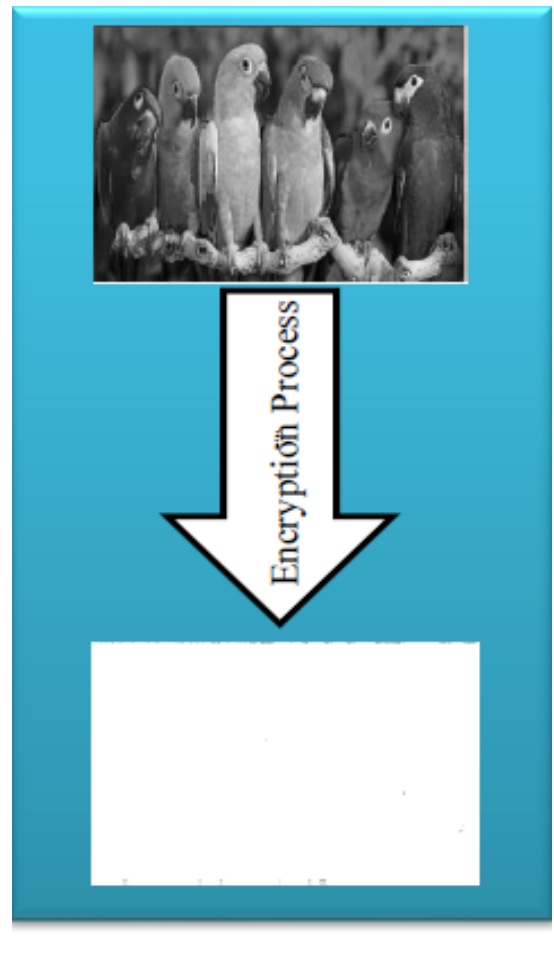

Fig.(5): Encryption of Gray Image (png)

5. **Conclusion**

In this article, both color and grayscale image of any size saved in Portable network graphics(PNG), Joint Photographic Experts group (jpg) can be encrypted & decrypted using a modification the ElGamal cryptosystem Algorithm. However, the ElGamal cryptosystem security is based on the difficulty of finding discrete logarithms modulo a large prime , this modification gives better security over images because breaking this cryptosystem depends on solving discrete logarithm problem to get the private key (**a**) and knowing X . Therefore, figuring the private keys **a** and X much more harder than figuring only **a**. Therefore, this study suggests a modification of ElGamal cryptosystem over a primitive root of a large prime. This modification is applied on image to give more secure cryptosystem. This modification can make the ElGamal cryptosystem is more immune against some attacks than before. That leads to an increase of the confidence in the security of using this modification .